# Determinants of sexual transmission of HV: implications for control

B. G. Williams

South African Centre for Epidemiological Modelling and Analysis, Stellenbosch, South Africa
Correspondence to BrianGerardWilliams@gmail.com

**Abstract**

The extent to which ART (anti-retroviral therapy) reduces HIV transmission has received attention in recent years. Using data on the relationship between transmission and viral load we show that transmission saturates at high viral loads. We fit a power-law model and an exponential converging to an asymptote. Using data on the viral load in HIV-positive people we show that ART is likely to reduce transmission by 91.6% (81.7%–96.2%) under the first and 99.5% (98.5%–99.8%) under the second model.

    The role of the acute phase in HIV transmission is still debated. High levels of transmission during the acute phase have been used to argue that failure to identify people in the acute phase of HIV may compromise the impact of treatment on preventing new infections and that having concurrent sexual partners during the acute phase is an important driver of the epidemic. We show that the acute phase probably accounts for less than 1% of overall transmission. We also show that even if a significant proportion of infections are transmitted during the acute phase, this will not compromise the impact of treatment on population levels of transmission given the constraint implied by the doubling time of the epidemic.

    This analysis leads to other relevant conclusions. First, it is likely that discordant-couple studies significantly underestimate the risk of infection. Second, attention should be paid to the variability in set point viral load which determines both the infectiousness of HIV-positive people and the variability in the susceptibility of HIV-negative people. Third, if ART drugs are in short supply those with the highest viral load should be given priority, others things including age, gender and opportunistic infections being equal, but to reduce transmission ART should be offered to all those with a viral load above about 10k/mm.[3]

## Introduction

Understanding the relationship between HIV viral load and transmission is essential if we are to understand the dynamics of HIV and, in particular, the impact of ART on HIV transmission. It is generally accepted that the risk of HIV-transmission increases as the plasma viral load increases but precise data that can be used to address this question are few, especially at high and low viral loads. Transmission at low viral loads determines the impact that anti-retroviral therapy has on transmission; transmission at high viral loads determines the importance of transmission during the acute phase of HIV infection; transmission at intermediate viral loads determines the rate at which HIV spreads through a population.

    Previous attempts to establish a relationship between transmission and viral load assume a power-law relationship. Here we present an alternative model in which transmission increases linearly with viral load when the viral load is low but converges to an asymptote when the viral load is high. We consider the implications of both models for the impact of ART on the reduction in viral load and of ART on HIV transmission.

    We use data on the viral load over time to establish the duration and magnitude of the viral load during the acute stage of infection as compared to the chronic stage and to estimate the proportion of transmission events that take place during the acute phase. Some have argued that the acute stage of HIV infection lasts for up to five months during which time the risk of infecting another person is up to 30 times greater than in the subsequent chronic stage of infection with important implications for the dynamics and control of HIV. We show that this is very unlikely to be the case and attempt to reconcile our conclusions with this point of view.

    Finally, we consider likely biases in current estimates of the risk of infection per sexual encounter, based on studies of discordant couples, and the use of viral load as a way of triaging patients for ART.

## Data

For the frequency distribution of viral load among people in a generalized HIV epidemic we use data from a cross-sectional survey of young men in Orange Farm, South Africa.[1]

Attempts to measure the relationship between viral load and transmission have either presented the risk per sexual contact[2,3] or the risk per unit time and have included studies of discordant couples[3-10] and vertical transmission from mothers to their children.[11] To determine the relationship between transmission and plasma viral load we use data from three studies. The first by Attia et al.[6] is a meta-analysis of 6 earlier studies, the second by Donnell et al.[12] is from a study of the impact of treating HSV-2 on HIV-1 transmission, the third by Lingappa et al.[13] is based on prospective data from HIV-1 sero-discordant couples in East and southern Africa. To determine the relationship between transmission and viral load in genital secretions we use data from a prospective study of heterosexual sero-discordant couples in Africa that considered male to female and female to male transmission separately.[14] These studies all measured the number of transmission events per year which depends on the frequency of sexual encounters.

## Methods

### Estimating the distribution of viral load

In all of the studies of the relationship between transmission and viral load the authors only publish the number of transmission events in fixed ranges of viral load and we need to estimate the mean viral load for each range. To do this we start from the data in Figure 1 and determine the Box-Cox transformation[15]

$$w = \frac{v^\lambda - 1}{\lambda}, \qquad 1$$

with $w$ the transformed value of the viral load, $v$.

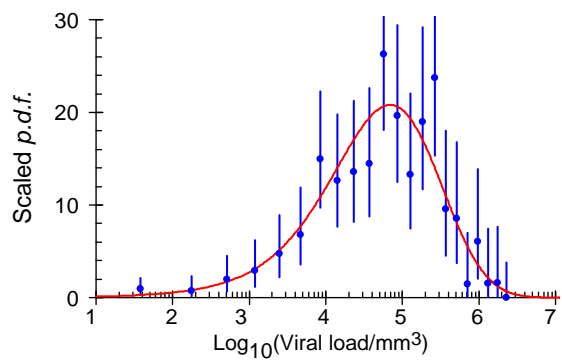

Figure 1. Measure and observed probability density function (*p.d.f.*) of viral load in young men from a cross-sectional survey carried out in Orange Farm, South Africa.[1]

The Box-Cox parameter is $\lambda = 2.58$ resulting in a normal distribution with a mean value of 20.8 and a standard deviation of 9.4. We then assume that in the study by Attia et al.[6] the density function is the same as in Orange Farm[1] while in the study by Donnell et al.[12] we keep the same value of the Box-Cox parameter but fit a cumulative normal distribution to the observed numbers of people in each range of the viral load[16] (see Appendix 1). For the Orange Farm study the mode of the viral load distribution is 95k virions/mL while in the Donnell et al.[12] study it is 29k virions/mL.

### Models of transmission as a function of viral load

We use two models to fit the data on transmission as a function of viral load. The first, used in all four of the publications cited,[6,12,13] assumes a power-law relationship between transmission and viral load; the second, which we develop here, assumes that at very low viral loads transmission increases linearly with viral load but converges exponentially to an asymptote when the viral load is very high. The equation for the power-law model is

$$T = \beta V^\sigma \qquad 2$$

and for the converging-exponential model is

$$T = \alpha \left(1 - e^{\rho V}\right). \qquad 3$$

so that at low values of the viral load

$$T \approx \alpha \rho V \qquad 4$$

## Results

### Distribution of viral load

The distribution of viral load before people start ART is variable but seldom published. For the data in Figure 1, measured in a cross-sectional survey of young men in Orange Farm, South Africa,[1] the mode of the distribution is at 53k/mL but with 95% of the distribution lying between 224/mL and 882k/mL, spanning more than three orders of magnitude. Those with a viral load in the region of $10^6$/mL will be more infectious than those with a viral load of 100/μL. But not only will treating those with the highest viral load have the greatest individual benefit, since survival decreases as viral load increases, but it will also have the greatest public health benefit by reducing transmission.

### Transmission as a function of plasma viral load

Figure 2 shows the data for the three studies of transmission as a function of plasma viral load fitted to the two models. For the data from Attia et al.[6] (Figure 2 A and B) the exponential model gives a statistically good fit ($p = 0.8$) while the power-law model does not ($p = 0.0008$). For the data from Donnell et al.[12] (Figure 2 C and D) and Lingappa et al.[13] (Figure 2 E and F) both models give statistically acceptable fits. On the basis of these data alone we cannot say with confidence that either model is better.

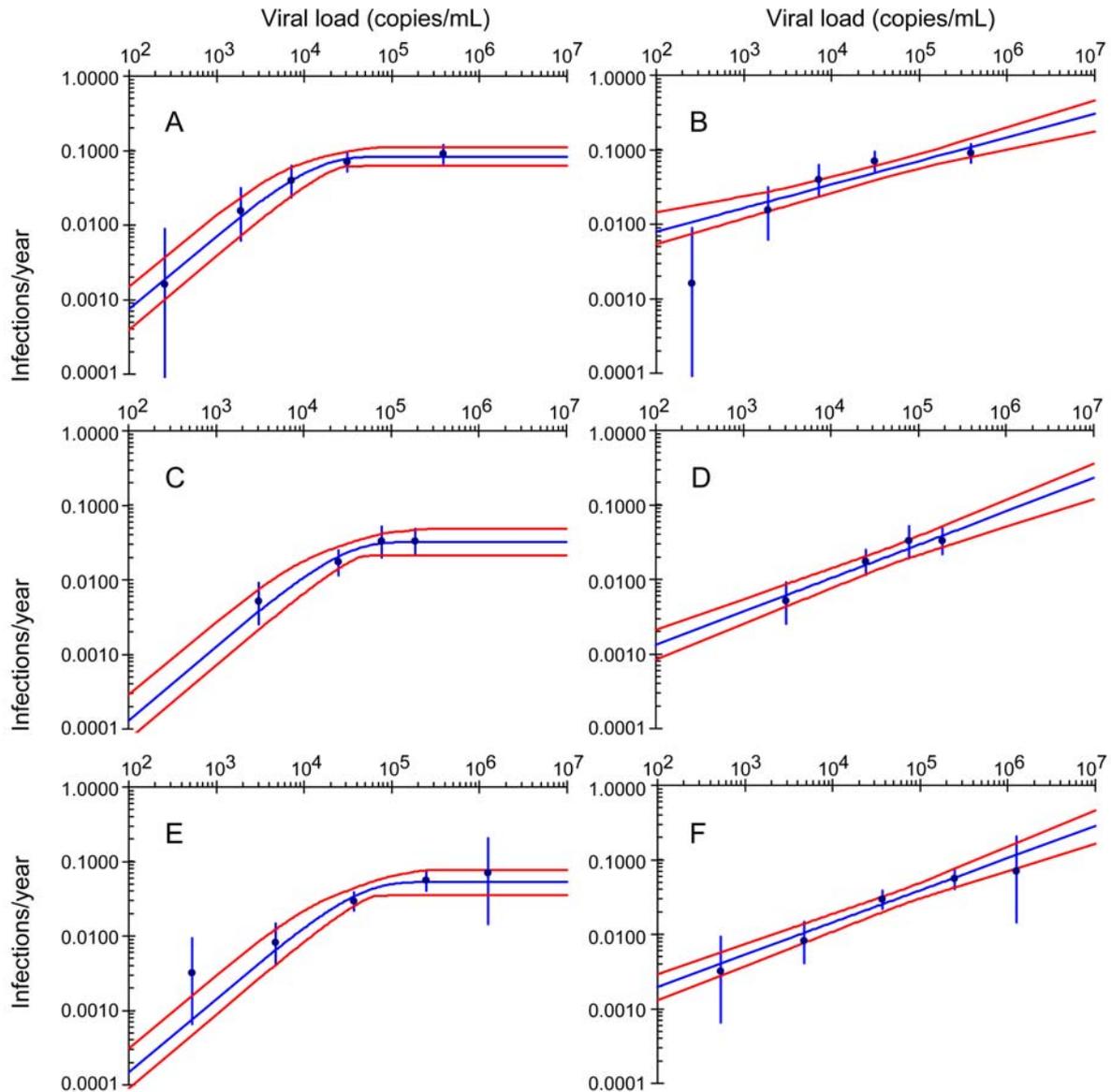

Figure 2. Two models fitted to data on the annual risk of infection in a discordant couple as a function of plasma viral load. Left: Converging-exponential (Equation 3). Right: Power-law (Equation 2). Data A and B: Attia et al.[6]; C and D: Donnell et al.[12]; E and F Lingappa et al.[13] The fitted parameters and the significance levels for the fits are A: $\alpha = 0.083$ (0.062–0.107) infections per year, $\rho = 2.23$ (2.14–2.31) mL/virion, $p = 0.80$; B: $\beta = 0.32$ (0.30–0.33), $\sigma = 0.0019$ (00.15–0.023), 0.0006; C: $\alpha = 0.032$ (0.024–0.042) infections per year, $\rho = 2.31$ (2.25–2.37) mL/virion, $p = 0.38$; D: $\beta = 0.45$ (0.42–0.47), $\sigma = 0.00017$ (0.00013–0.00022), 0.30; E: $\alpha = 0.053$ (0.042–0.066) infections per year, $\rho = 2.35$ (2.30–2.39) mL/virion, $p = 0.11$; F: $\beta = 0.43$ (0.41–0.45), $\sigma = 0.00027$ (0.00021–0.00034), $p = 0.39$.

Under the converging-exponential model, reducing the viral load from 50k/mL to 100/ml gives a relative risk for HIV-transmission of 0.53% (0.18% to 1.52%) corresponding to a 189 (66 to 556) fold reduction in transmission. Using the power-law model the corresponding relative risk is 8.4% (3.8% to 18.3%) corresponding to a 12 (5 to 26) fold reduction in transmission. Both reductions are substantial but their difference is significant and it will be important to establish which one is more likely to be correct since the extrapolation to low values of viral load depends critically on the choice of model. Both estimates are consistent with the results of a recent randomized controlled trial which showed that ART reduces transmission by 96% (73%–99%).[17]

Under the converging-exponential model, transmission saturates when the log-viral load is greater than 4.42 (4.40–4.44) and the viral load is 26.0 (24.8–27.3) k/mL. Under the power-law model, transmission increases as the viral load to the power of 0.32 (0.30–0.34), so that the increase is less than linear. In either case a biological explanation for the observation that transmission saturates at high values of viral load is needed.

The key features of the fits in Figure 2 are summarized in Table 1. To obtain an estimate of the reduction in transmission as viral load falls we compare the risk of infection at a viral load of 100/μL, which is achievable with anti-retroviral therapy,[18] with the risk at a viral load of 50k/mL, close to the mode of the distribution.

Table 1. Model 1 assumes that transmission increases linearly with viral load at low viral loads but converges exponentially to an asymptote. Model II assumes a power law relationship between transmission and viral load. *RR* is the risk of transmission at a viral load of 100/μL compared to that at $10^5$/mL. *SVL* is the saturation viral load (see text for details). *Power* is the power in the power-law relation between transmission and viral load. The data are plotted in Figure 2. The estimates in each of the first three rows are not significantly different ($p > 0.67$ in all cases) but the estimates in the bottom row are significantly over-dispersed ($p < 0.0001$). Mean values are therefore weighted means using the estimated errors for the first three rows and unweighted means, with the error calculated from the residuals, in the bottom row.

| Model | | Attia et al.[6] | Donnell et al.[12] | Lingappa et al.[13] | Mean |
|---|---|---|---|---|---|
| I | RR (%) | 0.91 (0.38–2.43) | 0.46 (0.20–1.37) | 0.37 (0.18–1.01) | 0.53 (0.18–1.52) |
|   | $Log_{10}$(SVL) | 4.04 (3.69–4.36) | 4.39 (4.12–4.65) | 4.56 (4.33–4.76) | 4.42 (4.40–4.44) |
| II | RR (%) | 14.05 (7.76–31.4) | 6.19 (3.08–13.0) | 6.87 (3.60–12.80) | 8.35 (3.80–18.3) |
|   | Power | 0.32 (0.30–0.33) | 0.45 (0.42–0.47) | 0.43 (0.41–0.45) | 0.32 (0.30–0.34) |

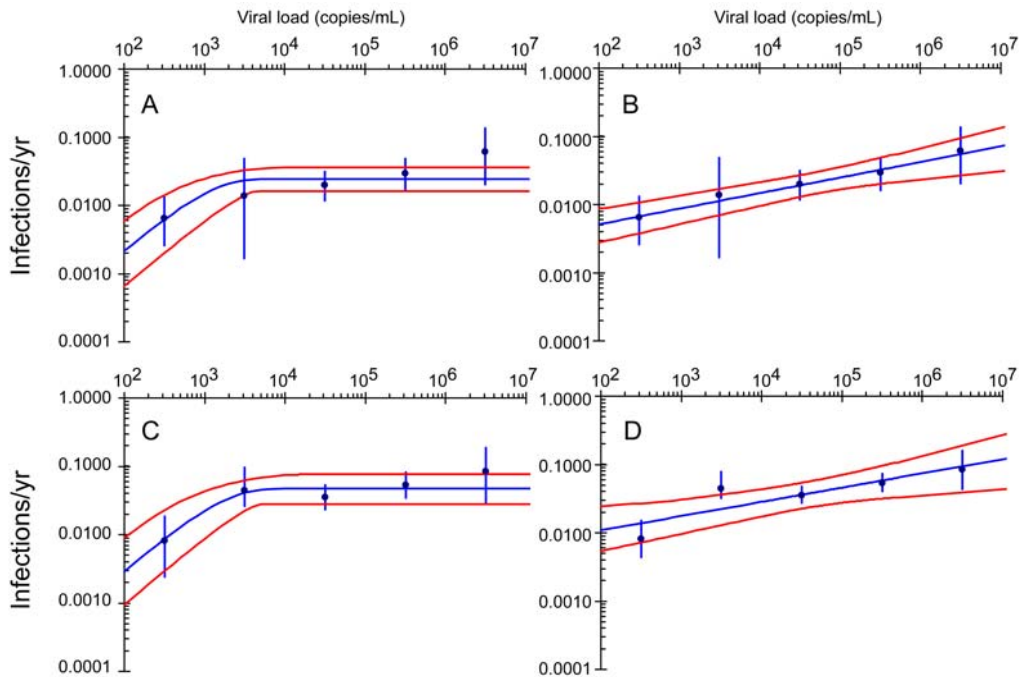

Figure 3. Two models fitted to the annual risk of infection in discordant couples as a function of viral load in genital secretions.[14] Left: converging-exponential (Equation 3). Right: power-law (Equation 2). A and B: female to male transmission; C and D: male to female transmission. A: $\alpha = 0.025$ (0.017–0.035) infections per year, $\rho = 1.55$ (1.28–1.78) mL/virion, $p = 0.16$; B: $\beta = 0.23$ (0.21–0.24), $\sigma = 0.0030$ (00.26–0.035), 0.96; C: $\alpha = 0.047$ (0.030–0.071) infections per year, $\rho = 1.62$ (1.36–1.82) mL/virion, $p = 0.55$; D: $\beta = 0.21$ (0.15–0.25), $\sigma = 0.0067$ (0.0042–0.0101), $p = 0.22$.

**Transmission as a function of genital viral load**

In determining HIV transmission, the viral load in genital secretions is likely to be more important than the viral load in plasma. Figure 3 shows the limited available data and the fitted curves for the relationship between transmission and viral load in genital secretions. Both models give acceptable fits to the data so that one cannot determine which model is better on the basis of the statistical fit

The data in Figure 3 are summarized in Table 2. The asymptotes are slightly, but not significantly, lower than the asymptotes in Figure 2 but the data in Figure 3 suggest that the saturation viral load in genital secretions is 197 (90–430) times, or about two orders of magnitude, less than in the plasma. Using the power-law model gives a correspondingly low value for the power in the fit. The reason for the saturation effect remains to be explained. These data are less precise than the data based on plasma viral load and reducing the genital viral load to 100/μL would reduce transmission by 92.6% (31.9%−99.2%) using Model I and by 84.7% (21%−97.0%) using Model II.

Table 2. Relative risk of transmission at high and low viral loads, the saturation viral load and the power-law relationship between viral load and transmission. Model 1 assumes that transmission increases linearly with viral load at low viral loads but converges exponentially to an asymptote. Model II assumes a power law relationship between transmission and viral load. *RR* gives the risk of transmission at a viral load of 10/μL compared to that at $10^5$/mL. *SVL* is the saturation viral load (see text for details). *Power* is the power in the power law relation between transmission and viral load. The estimates do not differ significantly; $p > 0.95$ in all cases. M: male; F: female.

| Model | | Baeten et al. F to M[14] | Baeten et al. M to F[14] | Mean |
|---|---|---|---|---|
| I | *RR* (%) | 8.72 (1.82–37.1) | 6.17 (1.24–32.4) | 7.44 (0.81–68.1) |
|   | $Log_{10}$(SVL) | 2.04 (1.57–2.57) | 2.20 (1.68–2.69) | 2.12 (1.51–2.98) |
| II | *RR* (%) | 14.31 (5.11–40.8) | 17.0 (5.07–74.9) | 15.25 (2.95–79.0) |
|    | Power | 0.23 (0.21–0.24) | 0.21 (0.15–0.25) | 0.23 (0.20–0.26) |

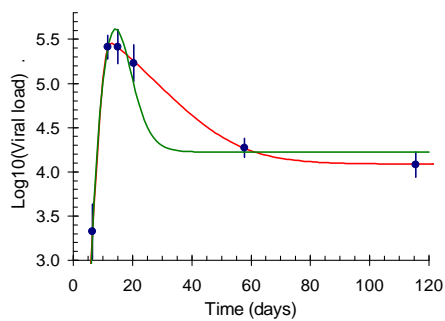

Figure 4. Viral load as a function of time[19] after infection fitted to an exponential converging to an asymptote that determines the peak viraemia followed by a logistic decline to the set point. The two curves represent the longest (25 days) and shortest (12 days) durations of the acute phase that are consistent with the data.

**Viral load during the acute phase**

After a person has been infected with HIV the viral load increases rapidly with a doubling time of 20.5 (18.2–23.4) hours.[19] Assuming that each infection is established from a single virion, it will take about 3.5 weeks to reach a concentration of $10^5$ virions per mL. in a person with 5 litres of blood. In order to fully characterize the acute phase, HIV negative people would have to be tested once or twice a week to ensure that the initial rise is accurately captured. An early study, based on data from newly infected plasma donors[19] is shown in Figure 4. Using the full-width at half-maximum above the set point (FWHM), on a log-scale, gives an acute phase duration of between 12 and 25 days, a peak viraemia of $10^{5.4}$ = 251k per mm$^3$ and a set point viraemia of $10^{4.1}$ = 13k per mm$^3$.

Preliminary, and only partial, results from an ongoing study have been reported by Robb.[20] People from high risk populations in Uganda, Kenya, Tanzania and Thailand are tested for HIV twice a week and as soon as they are found to be HIV positive they are followed up intensively for the first month and then for up to five years. A subset of the data, taken from a presentation,[20] are shown in Figure 5. Averaged over all of the data sets the FWHM is 17 ± 3 days, the peak viraemia is $10^{6.6±0.4}$ = 3.0 (1.6–10.0) million virions per mm$^3$ and the set point viraemia is $10^{4.5±0.5}$ = 32 (10–100) thousand virions per mm$^3$.

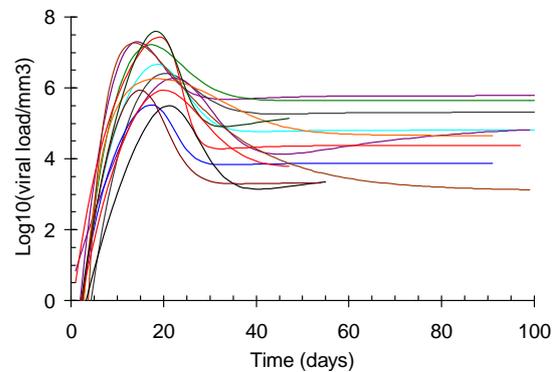

Figure 5. Data from the acute phase study of Robb.[20] The data have been extracted from a presentation. The fitted curves are exponentials converging to an asymptote followed by a logistic decline to a steady state

In another study Powers et al.[21] cite data from Pilcher et al.[22] for which the FWHM is 2.3 weeks and which Powers et al.[21] interpret as giving an acute phase duration of two weeks (Powers et al.,[21] their supporting information, Figure 1).

Taken together these data suggest that the acute phase lasts for about 2 to 3 weeks, that the $\log_{10}$ of the peak viraemia per mm$^3$ is 6.0 ± 1.0 while the $\log_{10}$ of the set point viraemia per mm$^3$ is 30 times lower at 4.5 ± 0.5.

**Transmission during the acute phase**

It has been suggested that transmission in the acute phase contributes a significant proportion of overall transmission.[21,23] Given the difficulty of finding people during the short acute phase this has important consequences for the impact of treatment on transmission.

We first consider the risk of infection during the acute as compared to the chronic phase of infection. Using the converging-exponential model (Figure 2 A, C and E) and assuming that the $\log_{10}$ of the set point viral load is 6.0 in the acute phase and 4.5 in the chronic phase, the risk of infection in the acute phase is between 1% and 2% higher than in the chronic phase, and less than 7% higher, with 95% certainty, for all three sets of data. Using the power-law model, with a power of 0.23 (Table 2) suggests that transmission in the acute phase is 2.2 (2.0–2.5) times higher than in the chronic phase. Given that the chronic phase lasts for an average of about 11 years, these data suggest that the proportion of transmission that takes place during the acute phase is between about 0.4% using the converging-exponential model, and 0.8% using the power-law model. In either case it is unlikely that the high viral load during the acute phase makes a significant difference to the rate of infection.

**Transmission among discordant couples**

We attempt to reconcile the observation that the relatively high viral load in the acute phase does not contribute significantly to overall transmission with claims that transmission among discordant couples is 7.2 (3–17) times higher in the first five months after infection than in the chronic stage.[24] In other analyses of the same data transmission was 26 (13–54) times higher in the first 2.9 months[25] or 13 times higher in the first 2.5 months[26] after infection than in the chronic stage. These two estimates are not themselves inconsistent, given the uncertainty in the original data and the assumptions concerning the duration of the acute phase. However, if the acute phase lasts for only two weeks, the infectivity during this short time would have to be of the order of 70 (30–170) times higher than during the chronic phase.

Three factors may help to reconcile these observations. The first is that, as Wawer et al.[24] note in their original paper, genital ulcer disease is significantly associated with the risk of HIV infection and the risk of HIV infection may increase by 6.0 (2.6–14.0) times in those recently infected with *Herpes simplex* (HSV) 1 as compared to those with an established HSV-1 infection or without HSV-1 infection.[27] Given the high incidence of HSV-1 in parts of Africa[28] it is not unlikely that people who are infected from outside the relationship will be infected at about the same time with both HIV-2 and HSV-1 with a corresponding increase in the likelihood that they will infect their sexual partners.

The second factor is that different people are more or less susceptible to HIV-infection.[29] Furthermore, there is evidence that young women are at especially high risk of HIV infection for both biological and social reasons and Wawer et al.[24] show that the risk of infection is 2.2 (1.2–4.0) times higher in those younger then 30 years of age as compared to those older than 30 years of age. In a cohort study those that are most susceptible to infection will be infected first leaving those that are at lower risk of infection to be infected later.

The third factor is that different people are more or less infectious to others. As shown in Figure 1 and Figure 5, the set-point viral load varies among people infected with HIV. Since those with high set-point viral loads will be more likely to infect their partners first the transmission probability is expected to fall over time spent in the cohort. To estimate the importance of this, we use the risk of transmission as a function of viral load given in Figure 2A and let survival in years, $S$, vary as the plasma viral load/mm$^3$ as in Appendix 4.

We let the viral load distribution in incident HIV cases be log-normally distributed with a mean value of the $\log_{10}$(viral load/mm$^3$) of 5.14 and a standard deviation of 0.78 (Appendix 4). We can then estimate the probability that the index case will remain alive and the susceptible partner will remain uninfected, as a function of time, and hence the average transmission rate as a function of time since the index case was infected given that the index case is alive and the partner is still uninfected. We also need to estimate the initial rate of transmission immediately after the index case is infected. To do this we note that the initial epidemic doubling time in heterosexual epidemics is typically about 15 months[30] so that the transmission rate, at the start of the epidemic, is about 0.8 per year. This initial transmission rate implies that after five months 28% of people in discordant relationships should be infected which is not significantly different from the estimate given by Wawer et al.[24] who found that 10/23 or 43% (23%–66%) of negative partners in discordant couples were infected in the first five months although this latter estimate does seem to be very high.

Proceeding in this way we are able to obtain the risk of transmission in discordant couples, as a function of time spent in the cohort, allowing for the fact that HIV-positive people with high viral load will both be more likely to infect their partners sooner and will be more likely to die sooner.

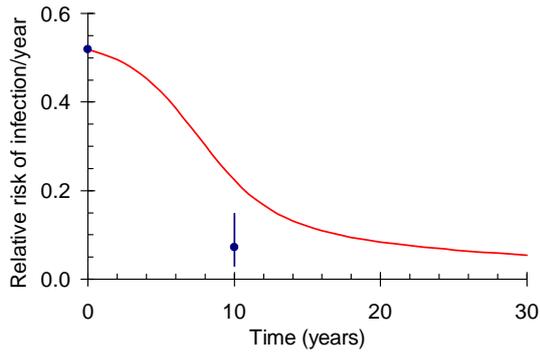

Figure 6. The risk of transmission in a discordant couple compared to the risk immediately after the index partner becomes infected as a function of time spent in the cohort allowing for the fact that those with high viral load are most infectious but will also die soonest.

Figure 6 suggests that after ten years transmission rates in discordant couples may have fallen by about 60% which is still significantly less than the decline in relative risk estimated by Wawer et al.[24] assuming that their 'prevalent cases' had been infected for an average of ten years.

Variability in the infectiousness and survival of people resulting from the variability in viral load may explain some, but not all, of the variability given in the study by Wawer et al.[24] It would be of interest to include variability in the susceptibility to infection, where such data are available, to see if high rates of transmission soon after infection with HIV can be used to explain the results of Wawer et al.,[24] without appeal to the acute phase, and also resolve the apparent paradox that assumed rates of transmission appear to be insufficient to sustain an epidemic of HIV.

### High rates of transmission in the acute phase*

This analysis suggests that the duration of the acute phase is too short and the increased risk of infection during the acute phase too small to have a significant impact on the overall dynamics of the epidemic. However, there are those that maintain that the duration of the acute phase may be several months and that during this time the risk of infection per sex act maybe increased by up to about 30 times.[21,23] Given that this could be the case it behoves us to consider the consequences that this would have for treatment as prevention.

The key point is this: one of the few directly observed parameters concerning the epidemiology of HIV is the initial doubling time which, in South Africa, is $1.25 \pm 0.25$.[30] Since the acute phase lasts for considerably less time than the chronic phase, the greater the relative risk of transmission in the acute phase the smaller must be the value of $R_0$ to maintain the same initial doubling time. Indeed, if we know the initial doubling time (the growth rate $r$ in Appendix 2, Equation 11), the relative risk of infection in the acute phase and each of the four chronic phases ($\lambda_i/\lambda_0$ in and 9), and the duration of each of the four stages ($1/\delta_i$ in Appendix 2, Equations 8 and 9) in Appendix 3, we can determine the value of $\lambda_0$ that gives the observed rate of increase of the prevalence at the start of the epidemic and hence the value of $R_0$ (Appendix 2, Equations 16).

Values of $R_0$, as a function of the relative risk of transmission during the acute phase and the duration of the acute phase in months, are given in Figure 7A. Without ART the value of $R_0$ is 5.8 (brown rectangle). With $RR = 2.1$ and $D_{AP} = 2$ mo. (green ellipse) $R_0$ falls to 5.4. With $RR = 8.3$ and $D_{AP} = 6$ mo. (red ellipse) $R_0$ falls to 3.0. With $RR = 26$ and $D_{AP} = 6$ mo. (blue ellipse) $R_0$ falls to 2.3. As expected the higher the rate of transmission during the acute phase the lower the value of $R_0$.

The boundaries of the ellipses indicate the uncertainty in the point estimates which are considerable. Assuming, as noted above, that in the Hollingsworth et al.[25] study the high values of $D_{AP}$ correspond to low values $RR$, and vice versa, we slant the corresponding confidence ellipse at an appropriate angle. This also serves to show that if we let $D_{AP} = 6$ mo. the estimates made by Hollingsworth et al.[25] and the Wawer et al.[24] are not significantly different.

It is important to note that testing people regularly at intervals of one year, say, is more efficient than testing people randomly but once a year on average. In the latter case some people are never tested and some people are tested more often than is strictly necessary. This is discussed further in Appendix 3. Figure 7B shows what happens if people are tested randomly but once a year on average. With $RR = 1$, the value of $R_0$ falls to 0.58 (brown rectangle). With $RR = 2.1$ and $D_{AP} = 2$ mo. $R_0$ falls to 0.61. With $RR = 8.3$ and $D_{AP} = 6$ mo. $R_0$ falls to 0.84 and with $RR = 26$ and $D_{AP} = 3$ mo. $R_0$ falls to 0.90. In all three cases $R_0$ still falls below 1 although with the two higher estimates of $RR$ it is only 10% to 20% below 1 which allows for little margin for error.

Figure 7C shows what happens if the average testing interval is reduced to six months. With $RR = 1$, $R_0$ falls to 0.29 (brown rectangle). With $RR = 2.1$ and $D_{AP} = 2$ mo. $R_0$ falls to 0.34. With $RR = 8.3$ and $D_{AP} = 6$ mo. $R_0$ falls to 0.61 and with $RR = 26$ and $D_{AP} = 3$ mo. $R_0$ falls to 0.75. Even in the worst case ($RR = 26$, $D_{AP} = 3$ mo.) $R_0$ is significantly less than 1.

---

\* For an earlier, but less complete, discussion of the issues raised in this section see: http://arxiv.org/abs/1105.2767.

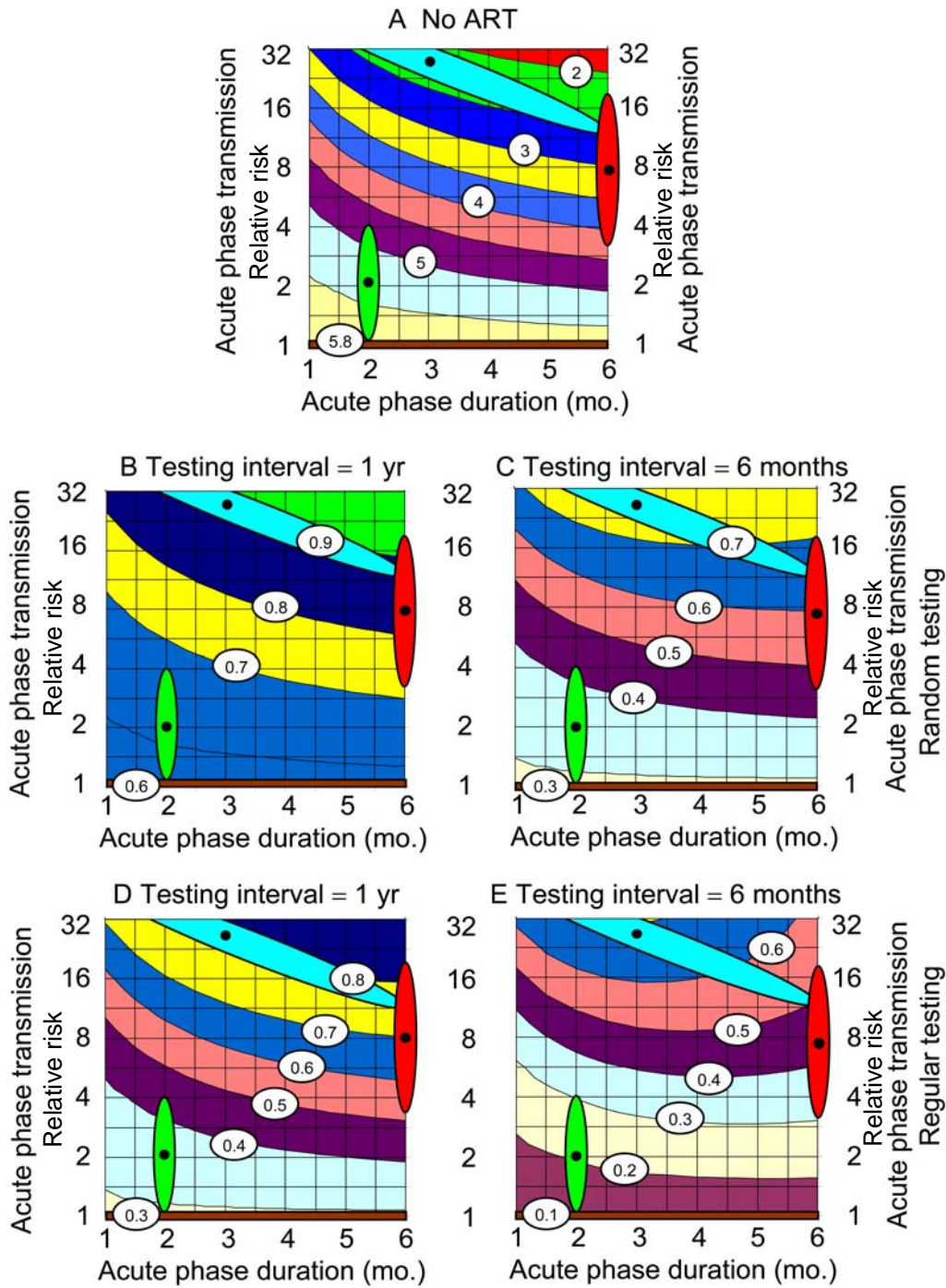

Figure 7. The value of $R_0$ for an epidemic with a doubling time of 15 months as a function of the duration of the acute phase and the relative transmission rate in the acute phase compared to the chronic phase. Colours correspond to different values of $R_0$ and the numbers in circles give the values on different contour lines. A without ART; B and C with random testing once a year or twice a year, on average; D and E with regular testing once a year or twice a year. Brown rectangle: $RR = 1$; green ellipse: $RR = 2.1$, $D_{AP} = 1$ mo.; ellipse $RR = 8.3$, $D_{AP} = 6$ mo.; blue ellipse: $RR = 26$, $D_{AP} = 3$ mo. The ellipses are 95% confidence limits of the point estimates.

Figure 7D shows what happens if people are tested regularly once a year. With $RR = 1$, the value of $R_0$ again falls to 0.29 (brown rectangle). With $RR = 2.1$ and $D_{AP} = 2$ mo. $R_0$ falls to 0.38. With $RR = 8.3$ and $D_{AP} = 6$ mo. $R_0$ falls to 0.70 and with $RR = 26$ and $D_{AP} = 3$ mo. $R_0$ falls to 0.82. Again, even in the worst case ($RR = 26$, $D_{AP} = 3$ mo.) $R_0$ is significantly less than 1.

Figure 7E shows what happens if people are tested regularly twice a year. With $RR = 1$, $R_0$ falls to 0.14 (brown rectangle). With $RR = 2.1$ and $D_{AP} = 2$ mo. $R_0$ falls to 0.21. With $RR = 8.3$ and $D_{AP} = 6$ mo. $R_0$ falls to 0.45 and with $RR = 26$ and $D_{AP} = 3$ mo. $R_0$ falls to 0.68. Even in the worst case ($RR = 26$, $D_{AP} = 3$ mo.) $R_0$ is again significantly less than 1.

We have three estimates of $RR$, the relative risk of infection, and $D_{AP}$, the duration of the acute phase ranging from 2.1 over 2 months to 26.2 over 3 months giving values of $R_0$ ranging from 5.8 to 2.3. The high estimates for $RR$ may be over-estimates. They are both based on the data from Rakai[24] and discordant couple studies in which one partner is already infected will select against those couples who are most infectious and therefore no longer sero-discordant, as discussed above. Furthermore, the high values of $RR$ with long values of $D_{AP}$ imply values of $R_0 \approx 2$ which seems unlikely; if this were the case HIV should be relatively easy to eliminate through minor changes in behaviour and the epidemic should be much less stable.

Even if we adopt the most pessimistic view and assume that the relative risk of infection is 26 times higher during an acute phase that lasts for 3 months annual testing and immediate treatment has the potential to reduce $R_0$ to less than 1 and with any further contribution to prevention will guarantee elimination in the long term. Testing people regularly, on an annual basis, is more effective than random testing because under random testing some people will be tested very frequently, which is not necessary, while others will be tested very infrequently which is not ideal. With regular testing even the most pessimistic view reduces $R_0$ to 0.82 and will probably lead to elimination. As expected, testing people twice a year reduces $R_0$ even further and under all assumptions about the acute phase would guarantee elimination.

## Using viral load to triage patients

Using CD4+ cell counts to decide which patients are in greatest need of ART is flawed for a number of reasons. First, CD4+ cell counts in HIV-negative people vary widely. In Orange Farm, South Africa, 95% of HIV-negative people have CD4+ cell counts in the range 380/μL to 1550/μL.[1] About 10% of the population have CD4+ cell counts below 500/μL before they are infected with HIV and given that CD4+ cell counts drop by about 25% immediately after sero-conversion 33% of people will have CD4+ cell counts below 500/μL within two weeks of being infected with HIV.

Second, CD4+ cell counts vary widely among populations. In Orange Farm, South Africa, the median CD4+ cell count in HIV-negative people is 1115/μL[1] while in Botswana it is 599/μL.[31]

Thirdly, if we consider a country such as Zimbabwe, where the prevalence of infection first rose and then fell rapidly, we can show that in 1985 about 10% of those infected had a CD4+ cell count below 350/μL but by 2005 about 52% had a CD4+ cell count below 350/μL.

Finally, at very low CD4+ cell counts people are likely to have easily diagnosed AIDS defining illnesses and at high CD4+ cell counts mortality rates are relatively insensitive to CD4+ cell counts.[32]

Since viral load is a much better prognostic indicator than CD4+ cell counts of both survival and infectiousness, some scientists have suggested using viral load tests to triage patients for ART and trials are being developed in which people will be started on ART if their viral load is less than 50k/mL.[33]

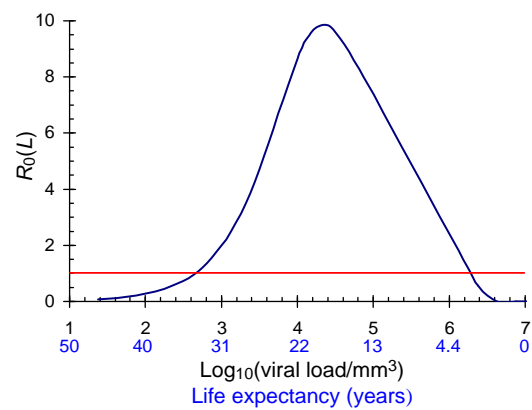

Figure 8. $R_0(L)$, the value of $R_0$ as a function of set point viral load. Blue numbers on the horizontal axis give the life expectancy for each viral load. For this model $\langle R_0(L) \rangle = 6.0$.

From the distribution of viral load at the sent point (Appendix 4), transmission as a function of viral load (Figure 2A), and the life expectancy as a function of viral load (Appendix 4, Figure 10) we can calculate the value of $R_0(L)$, the number of secondary infections for each primary infection as a function of viral load. This is shown in Figure 8. Those with a set-point viral load of $10^{4.4} = 25k/mm^3$ have the potential to contribute most to onward transmission but this will take place over about 20 years. Furthermore, only those with a set point viral load between $10^{2.7} = 501/mm^3$ and $10^{6.3} = 2M/mm^3$ have a value of $R_0(L) > 1$.

We can also plot the cumulative proportion of infections as a function of viral load as shown in Figure 9. We see that only 23% of infections are contributed by those with a viral load greater than $100k/mm^3$ and 95% of all infections are contributed by those with a viral load between $10^{2.9} = 794/mm^3$ and $10^{5.7} = 502k/mm^3$.

The *caveat* associated with this analysis is that at low viral loads the secondary infections will only occur over a long period of time. From the point of view of the individual patients one should give priority to those with the highest viral load, other things being equal. But to significantly reduce transmission it will also be necessary to ensure that all those with a viral load above about 100k/mm$^3$ are offered ART since it is only below this level that infectiousness starts to decline rapidly with viral load (Table 1).

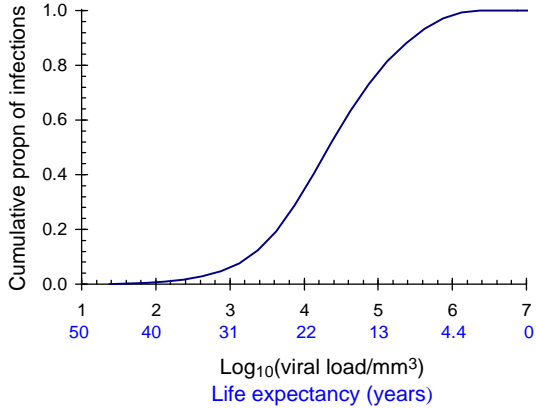

Figure 9. $R_0(L)$, the value of $R_0$ as a function of set point viral load. Blue numbers on the horizontal axis give the life expectancy for each viral load. For this model $\langle R_0(L) \rangle = 6.0$.

## Conclusions

There is strong evidence that transmission saturates, as viral load increases, at about $10^4$ to $10^5$ virions per mm$^3$ and this demands a biological explanation. This implies that particularly high viral loads during the acute phase are unlikely to be important.

The available data suggest that people who are fully compliant with ART can reduce their infectiousness to others by at least 99.4% (98.5% to 99.8%), consistent with the recent HPTN 052 study which gave a reduction of 96% (73%–99%).[34]

There is evidence that the duration of the acute phase is of the order of 2 to 3 weeks or about 0.3% of the average disease duration. This makes it even less likely that the acute phase contributes significantly to transmission; our best estimate is that the acute phase accounts for no more than 1% of total transmission.

In the unlikely event that the rate of transmission during the acute phase is about ten times higher than in the chronic phases and if the acute phase lasts, unrealistically, for two months, constraining the initial rate of increase of HIV-prevalence to between one and two years implies a value of $R_0$ that is not much greater than 1. Even these extreme assumptions would not jeopardize the impact of treatment on transmission.

This analysis suggests that previous estimates of the risk of infection per unprotected sexual encounter, based on studies of discordant couples, considerably underestimate the risk of infection with important implications for interpreting models of HIV transmission.

What is needed in models of the impact of ART on the dynamics of HIV is a) consideration of the variability in the set point viral load and b) consideration of the variability in the susceptibility of individual people, and the implications that both of these have for the infectiousness and the survival of individual people.

Finally we note that, especially in relation to the acute phase, having concurrent partners is very unlikely to affect transmission in agreement with the conclusions of others.[35] Of considerably greater interest than the currently sterile debate concerning concurrency[36,37] would be a thoughtful and nuanced analysis of the implications of sexual network structures on the epidemic of HIV noting that concurrency, as currently and rather poorly defined, constitutes a small aspect of a bigger, and much more interesting, question.

## Appendix 1 Transforming and fitting the density function of viral load

Let $f(v)$ be the probability density function of $v$, the logarithm of the viral load to the base ten. We apply a Box-Cox transform to $v$ so that the transformed variable $w$ is

$$w = \frac{v^\lambda - 1}{\lambda} \qquad 5$$

where $\overline{v}$ is the geometric mean of the data. The density function of $w$ is then

$$g(w) = f(v)\frac{dv}{dw} \qquad 6$$

where

$$\frac{dv}{dw} = v^{\lambda-1} \qquad 7$$

We first transform the data using Equation 5 and vary the parameter $\lambda$ to get the best fit. We then fit a normal distribution to the transformed data and carry out the reverse transform and this give the curve shown in Figure 1.

For the data from Orange Farm, South Africa[11] the Box-Cox parameter is $\lambda = 2.580$ and the transformed data is normally distributed with a mean of 20.8 and a standard deviation of 9.4.

To determine the density function for the data in the study by Attia *et al.*[6] we assume that the distribution is the same as for the Orange Farm data. To determine the density function for the data in the study by Donnell *et al.*[12] and Lingappa *et al.*[13] we apply the Box-Cox parameter for Orange Farm to these data and then fit the observed cumulative distribution function to a normal cumulative distribution function. In both cases, once we have the density function we can then work out the mean value of the viral load within each range as given in the papers.

## Appendix 2 $R_0$ and the growth rate for an $n$-stage model

We want to introduce an acute phase but also keep four chronic stages in order to ensure that the survival distribution approximates the observed Weibull distribution with a shape parameter of 2.25.[38] We let $\lambda_i$ be infectivity of people in stage $i$ of infection and $\delta_i$ be the rate at which people in stage $i$ progress to the next stage of infection. Let $P_i$ be the proportion of people in stage $i$ ($i = 0$ for those that are uninfected) and assume that the death rate, when people leave from the last stage, match the recruitment rate so that we are dealing with proportions. The equations for the model are then (keeping the total population constant)

$$P_0^\bullet = \lambda_n P_n - P_0 \sum_{i=1}^{n} \lambda_i P_i \qquad 8$$

$$P_1^\bullet = P_0 \sum_{i=1}^{n} \lambda_i P_i - \delta_1 P_1 \qquad 9$$

$$P_i^\bullet = \delta_{j-1} P_{j-1} - \delta_j P_j, \quad j = 2 \text{ to } n \qquad 10$$

At the start of the epidemic $P_i \approx 1$ and, once the overall prevalence is increasing at a constant rate $\rho$ the prevalence of those in each infected stage will be increasing at the same rate $\rho$ so that

$$P_i^\bullet = r P_i \quad i = 1 \text{ to } n \qquad 11$$

and

$$\frac{P_{j+1}}{P_j} = \frac{\delta_j}{\rho + \delta_{j+1}} \quad i = 1 \text{ to } n \qquad 12$$

From Equations 9 and 11

$$\sum_{i=1}^{n} \lambda_i P_i = (\rho + \delta) P_1 \qquad 13$$

Expanding Equation 12 gives

$$\frac{\lambda_1}{\rho + \delta_1} + \frac{\lambda_2}{\rho + \delta_2} \frac{P_2}{P_1} + \ldots + \frac{\lambda_n}{\rho + \delta_n} \frac{P_n}{P_1} = 1 \qquad 14$$

and substituting Equation 12 in Equation 13 we get

$$\sum_{i=1}^{n} \lambda_i \prod_{j=1}^{i} \frac{\delta_{i-1}}{\rho + \delta_i} = 1 \qquad 15$$

with $\delta_0 = 1$.

We now set duration of each stage, $\delta_i$, the risk of infection in each stage, $\lambda_i$, relative to the first stage so that $\lambda_0$ will be allowed to vary but the rest are then determined, and we set $\rho$ to the observed rate of increase in prevalence at the start of the epidemic. We than vary $\lambda_0$ to find the value that satisfies Equation15. We then calculate $R_0$ directly as

$$R_0 = \sum_{i=1}^{n} \frac{\lambda_i}{\delta_i} \qquad 16$$

## Appendix 3 Random versus regular testing

Let the relative risk of infection vary with time since infection as $RR(t)$ Then under random testing at a rate $\rho$ year, the reduction in the overall transmission will be

$$R = \frac{\int_0^\infty e^{-\rho t} RR(t) dt}{\int_0^\infty RR(t) dt} \qquad 17$$

while under regular testing at an interval of $\tau$ years the reduction in overall transmission will be

$$R = \frac{\int_0^\tau \left(1 - \frac{t}{\tau}\right) RR(t)}{\int_0^\infty RR(t)} \qquad 18$$

Since we have estimates of the relative risk of transmission for different stages of infection, given that a person is alive, we approximate $RR(t)$ with an appropriate step function.

## Appendix 4. Viral load and survival

We assume that survival, after infection with HIV, is determined mainly by the set-point viral load and we wish to establish the relationship between viral load and survival. To do this we use the data on the distribution of viral load measure in Orange Farm, South Africa (Figure 1) and the survival distribution for people aged 15 to 24 from the Cascade Study[39] where the observed survival distribution is Weibull with a median value of 12 years and a shape parameter of 2.25.[38] We first determine the cumulative probability that a person has a given viral load from the probability density function in Figure 1. We then determine the survival corresponding to this cumulative probability from the Weibull survival function noting that this can be done analytically. There is a further adjustment that needs to be made. The data in Figure 1 give the proportion of prevalent cases as a function of viral load measured in a single cross-sectional survey while we need the distribution of the set-point viral load which is the incidence of prevalent cases. To do this we divide the prevalence data in Figure 1 by the survival and redo the calculation. Since this give a new survival function we then iterate to convergence. This gives the results shown in Figure 10.

The distribution of set-point viral loads, calculated in this way, is then very close to a log-normal distribution, as suggested by Fraser et al.,[40] with a mean value of the $\log_{10}$(viral load/mm3) of 5.14 and a standard deviation of 0.78.

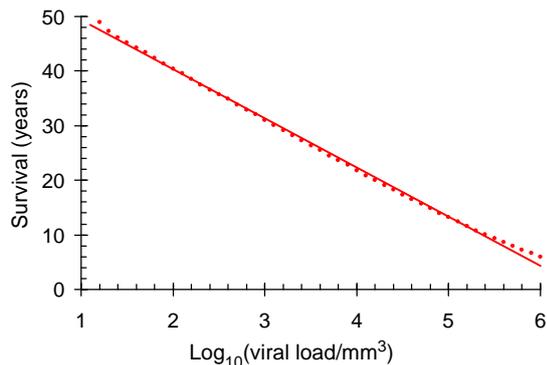

Figure 10. Survival ($S$) in years plotted against the logarithm of the viral load ($L$). A. The fitted line is $S = 58.4 - 9.00L$.

As a check on this relationship we compare the relationship implied by the data in Figure 10 with the direct data on survival, pre-ART, given by Arnaout et al.[41] Fitting a log-linear model to these data[41] gives a slope of $-6.7 \pm 2.7$ so that the slopes are not significantly different. However, the survival times are much shorted in the data from Arnaout et al.[41] The mean value of the $\log_{10}$(viral load/mm$^3$) is 4.7 at which point the survival is about 7 years whereas the corresponding survival in Figure 10 gives a survival time of about 16 years. The slope of the lines is similar but the intercept differs by about 2 years. We note also that in the study by Mellors et al.[42] mean value of the $\log_{10}$(viral load/mm$^3$) was 4.1 which is still lower than the value in the study by Arnaout et al.[41] These differences may well be due to differences in the populations studied.

## Acknowledgements

I thank an unknown friend of Victor Dukay for questioning my assumptions and leading me to carry out a more thorough analysis of the relationship between viral load and transmission. I thank John Hargrove who is always funny and supportive and who has argued, for many years, that discordant couple studies greatly underestimate per sex act transmission.

## References

1. Williams BG, Korenromp EL, Gouws E, Schmid GP, Auvert B, Dye C. HIV Infection, antiretroviral therapy, and CD4+ cell count distributions in African populations. J Infect Dis. 2006; **194**(10): 1450-8.

2. Peterman TA, Stoneburner RL, Allen JR, Jaffe HW, Curran JW. Risk of human immunodeficiency virus transmission from heterosexual adults with transfusion-associated infections. J Am Med Assoc. 1988; **259**(1): 55-8.

3. Quinn TC, Wawer MJ, Sewankambo N, Serwadda D, Li C, Wabwire-Mangen F, et al. Viral load and heterosexual transmission of human immunodeficiency virus type 1. Rakai Project Study Group. N Engl J Med. 2000; **342**(13): 921-9.

4. Tovanabutra S, Robison V, Wongtrakul J, Sennum S, Suriyanon V, Kingkeow D, et al. Male viral load and heterosexual transmission of HIV-1 subtype E in northern Thailand. J Acquir Immune Defic Syndr. 2002; **29**(3): 275-83.

5. Fideli US, Allen SA, Musonda R, Trask S, Hahn BH, Weiss H, et al. Virologic and immunologic determinants of heterosexual transmission of human immunodeficiency virus type 1 in Africa. AIDS Res Hum Retroviruses. 2001; **17**(10): 901-10.

6. Attia S, Egger M, Muller M, Zwahlen M, Low N. Sexual transmission of HIV according to viral load and antiretroviral therapy: systematic review and meta-analysis. AIDS. 2009; **23**(11): 1397-404.

7. Melo MG, Santos BR, De Cassia Lira R, Varella IS, Turella ML, Rocha TM, et al. Sexual transmission of HIV-1 among serodiscordant couples in Porto Alegre, southern Brazil. Sex Transm Dis. 2008; **35**(11): 912-5.

8. Castilla J, Del Romero J, Hernando V, Marincovich B, Garcia S, Rodriguez C. Effectiveness of highly active antiretroviral therapy in reducing heterosexual transmission of HIV. J Acquir Immune Defic Syndr Hum Retrovirol. 2005; **40**(1): 96-101.

9. Ragni MV, Faruki H, Kingsley LA. Heterosexual HIV-1 transmission and viral load in hemophilic patients. J Acquir Immune Defic Syndr Hum Retrovirol. 1998; **17**(1): 42-5.

10. Operskalski EA, Stram DO, Busch MP, Huang W, Harris M, Dietrich SL, et al. Role of viral load in heterosexual transmission of human immunodeficiency virus type 1 by blood transfusion recipients. Transfusion Safety Study Group. Am J Epidemiol. 1997; **146**(8): 655-61.

11. Garcia PM, Kalish LA, Pitt J, Minkoff H, Quinn TC, Burchett SK, et al. Maternal levels of plasma human immunodeficiency virus type 1 RNA and the risk of perinatal transmission. Women and Infants Transmission Study Group. N Engl J Med. 1999; **341**(6): 394-402.

12. Donnell D, Baeten JM, Kiarie J, Thomas KK, Stevens W, Cohen CR, *et al.* Heterosexual HIV-1 transmission after initiation of antiretroviral therapy: a prospective cohort analysis. Lancet. 2010; **375**(9731): 2092-8.

13. Lingappa JR, Hughes JP, Wang RS, Baeten JM, Celum C, Gray GE, *et al.* Estimating the impact of plasma HIV-1 RNA reductions on heterosexual HIV-1 transmission risk. PLoS ONE. 2010; **5**(9): e12598.

14. Baeten JM, Kahle E, Lingappa JR, Coombs RW, Delany-Moretlwe S, Nakku-Joloba E, *et al.* Genital HIV-1 RNA Predicts Risk of Heterosexual HIV-1 Transmission. Science Translational Medicine. 2011; **3**(77): 77ra29.

15. Box GEP, Cox DR. An analysis of transformations. J Roy Stat Soc B. 1964; **26**: 211-46.

16. Williams BG, Cutts FT, Dye C. Measles vaccination policy. Epidemiol Infect. 1995; **115**(3): 603-21.

17. Cohen MS, Chen YQ, McCauley M, Gamble T, Hosseinipour MC, Kumarasamy N, *et al.* Prevention of HIV-1 infection with early antiretroviral therapy. The New England journal of medicine. 2011; **365**(6): 493-505.

18. Palmer S, Maldarelli F, Wiegand A, Bernstein B, Hanna GJ, Brun SC, *et al.* Low-level viremia persists for at least 7 years in patients on suppressive antiretroviral therapy. Proc Nat Acad Sc USA. 2008; **105**(10): 3879-84.

19. Fiebig EW, Wright DJ, Rawal BD, Garrett PE, Schumacher RT, Peddada L, *et al.* Dynamics of HIV viremia and antibody seroconversion in plasma donors: implications for diagnosis and staging of primary HIV infection. AIDS. 2003; **17**(13): 1871-9.

20. Robb M. RV217: The early capture HIV cohort study (ECHO). Propsective identification of HIV infection prior to acute viraemia among high-risk populations. Poster Abstract 404. Presented at the Keystone Symposium. Whistler, British Columbia, Canada; 2011.

21. Powers KA, Ghani AC, Miller WC, Hoffman IF, Pettifor AE, Kamanga G, *et al.* The role of acute and early HIV infection in the spread of HIV and implications for transmission prevention strategies in Lilongwe, Malawi: a modelling study. The Lancet. 2011; **378**(9787): 256-68.

22. Pilcher CD, Joaki G, Hoffman IF, Martinson FE, Mapanje C, Stewart PW, *et al.* Amplified transmission of HIV-1: comparison of HIV-1 concentrations in semen and blood during acute and chronic infection. AIDS. 2007; **21**(13): 1723-30.

23. Cohen MS, Shaw GM, McMichael A, J., Haynes BF. Acute HIV-1 Infection. N Engl J Med. 2011; **364**(20): 1943-54.

24. Wawer MJ, Gray RH, Sewankambo NK, Serwadda D, Li X, Laeyendecker O, *et al.* Rates of HIV-1 Transmission per Coital Act, by Stage of HIV-1 Infection, in Rakai, Uganda. J Infect Dis. 2005; **191**(9): 1403-9.

25. Hollingsworth TD, Anderson RM, Fraser C. HIV-1 Transmission, by Stage of Infection. J Infect Dis. 2008; **198**(5): 687-93.

26. Abu-Raddad L, Longini IM, Jr. No HIV stage is dominant in driving the HIV epidemic in sub-Saharan Africa. AIDS. 2008; **22**: 1055–61.

27. Ramjee G, Williams B, Gouws E, Van Dyck E, Deken BD, Karim SA. The impact of incident and prevalent *Herpes Simplex* Virus-2 infection on the incidence of HIV-1 infection among commercial sex workers in South Africa. J Acquir Immune Defic Syndr Hum Retrovirol. 2005; **39**(3): 333-9.

28. Auvert B, Ballard R, Campbell C, Carael M, Carton M, Fehler G, *et al.* HIV infection among youth in a South African mining town is associated with *Herpes Simplex* virus-2 seropositivity and sexual behaviour. AIDS. 2001; **15**(7): 885-98.

29. Plummer FA, Ball TB, Kimani J, Fowke KR. Resistance to HIV-1 infection among highly exposed sex workers in Nairobi: what mediates protection and why does it develop? Immunol Lett. 1999; **66**(1-3): 27-34.

30. Williams BG, Gouws E. The epidemiology of human immunodeficiency virus in South Africa. Philos Trans R Soc Lond B Biol Sci. 2001; **356**(1411): 1077-86.

31. Bussmann H, Wester CW, Masupu KV, Peter T, Gaolekwe SM, Kim S, *et al.* Low CD4+ T-lymphocyte values in human immunodeficiency virus-negative adults in Botswana. Clin Diagn Lab Immunol. 2004; **11**(5): 930-5.

32. Williams BG, Hargrove JW, Humphrey JH. The benefits of early treatment for HIV. AIDS. 2010; **24**(11): 1790-1.

33. Lockman S. Test-and-Treat Strategy for HIV in Resource-Limited Settings. MedScape Today 2010 [cited; Available from: http://www.medscape.com/viewarticle/727536

34. Cohen MS, Chen YQ, McCauley M, Gamble T, Hosseinipour MC, Kumarasamy N, *et al.* Prevention of HIV-1 Infection with Early Antiretroviral Therapy. N Engl J Med. 2011; **July 18**.

35. Sawers L, Stillwaggon E. Concurrent sexual partnerships do not explain the HIV epidemics in Africa: a systematic review of the evidence. J Int AIDS Soc. 2010; **13**: 34.

36. Epstein H, Morris M. Concurrent partnerships and HIV: an inconvenient truth. J Int AIDS Soc. 2011; **14**: 13.

37. Padian NS, Manian S. The concurrency debate: time to put it to rest. The Lancet. 2011; **378**(9787): 203-4.

38. Williams BG, Granich R, Chauhan LS, Dharmshaktu NS, Dye C. The impact of HIV/AIDS on the control of tuberculosis in India. Proc Nat Acad Sc USA. 2005; **102**(27): 9619-24.

39. CASCADE Collaboration. Time from HIV-1 seroconversion to AIDS and death before widespread use of highly-active anti-retroviral therapy. A collaborative analysis. Lancet. 2000; **355**: 1131-7.

40. Fraser C, Ferguson NM, de Wolf F, Anderson RM. The role of antigenic stimulation and cytotoxic T cell activity in regulating the long-term immunopathogenesis of HIV: mechanisms and clinical implications. Proc R Soc London, Ser B. 2001; **268**(1481): 2085-95.

41. Arnaout RA, Lloyd AL, O'Brien TR, Goedert JJ, Leonard JM, Nowak MA. A simple relationship between viral load and survival time in HIV-1 infection. Proc Nat Acad Sc USA. 1999; **96**(20): 11549-53.

42. Mellors JW, Rinaldo CR, Jr., Gupta P, White RM, Todd JA, Kingsley LA. Prognosis in HIV-1 infection predicted by the quantity of virus in plasma. Science. 1996; **272**(5265): 1167-70.